\documentclass[preprint,authoryear,12pt]{elsarticle}

\usepackage{graphicx}
\usepackage{amssymb}
\usepackage{dcolumn}
\usepackage{bm}

\usepackage{color}

\usepackage[active]{srcltx} 
\usepackage{float} 

\newcommand{\nl}{u_{nl}}
\newcommand{\bk}{u_{bk}}

\newcommand{\dgs}{\delta_G^{I}}
\newcommand{\dge}{\delta_G^{II}}

\newcommand{\dd}{\delta}

\newcommand{\rw}{\rightarrow}

\newcommand{\mb}[1]{\mathbf {\boldsymbol  #1 }}

\journal{International Journal of Solids and Structures}

\begin{document}

\begin{frontmatter}



\title{Fatigue crack propagation in
a quasi one-dimensional elasto-plastic model}

\author{Tom\'as M. Guozden}
\ead{tguozden@cab.cnea.gov.ar}
\author{Eduardo A. Jagla}
\ead{jagla@cab.cnea.gov.ar}

\address{Centro At\'omico Bariloche and Instituto Balseiro, Comisi\'on Nacional de Energ\'{\i}a At\'omica, (8400) Bariloche, Argentina}



\begin{abstract}
Fatigue crack advance induced
by the application of cyclic quasistatic loads is investigated both numerically and analytically using a lattice spring model.
The system has a quasi-one-dimensional geometry, and consists in two symmetrical chains that are pulled apart, thus breaking springs which connect them, 
and producing the advance of a crack.
Quasistatic crack advance occurs as a consequence of the plasticity included in the
springs which form the chains, and that implies a history dependent stress-strain curve for each spring.
The continuous limit of the model allows a detailed analytical treatment
that gives physical insight of the propagation mechanism.
This simple model captures key features that cause well known phenomenology in fatigue crack propagation, in particular
a Paris-like law of crack advance under cyclic loading, and the overload retardation effect.

\end{abstract}

\end{frontmatter}

\section{Introduction}

Crack propagation usually occurs in one of two qualitatively different forms. 
In dynamic fracture \citep{freund}, propagation is very rapid, typically with velocities that are an 
appreciable fraction of the sound velocity in the material. Dynamic propagation occurs when some 
threshold load is exceeded. When this occurs, it becomes energetically favorable for the crack to advance. 
Once propagating, dynamical effects limit the maximum attainable crack velocity 
to some fraction of the sound velocity. Velocities in the range of 30-to-60 $\%$ of the sound velocity are
typically observed \citep{fineberg.99}.

Even if the threshold load for dynamical propagation is not reached, there is still the possibility of what is called
sub-critical crack growth. This kind of slow propagation can occur through different mechanisms. When the external load is kept constant, two different mechanism may lead to crack growth: creep and stress corrosion. Creep occurs at elevated temperatures, by degradation of the material ahead of the crack tip and corrosion occurs by a material-environment interaction close to the crack tip. Both of them are strongly time dependent.

Another class of sub-critical crack growth, is fatigue. More precisely, cyclic fatigue, which is the main focus of this paper.
It occurs when the external load has a cyclic dependence in time.
We limit the analysis to the cases where the external load changes very slowly in time, and so dynamical effects can be ignored altogether. We also
assume that a seed crack already exists in the system at the beginning of the process, as we do not intend to analize the problem of crack nucleation.
A necessary condition for an applied cyclic stress to produce crack advance is to reach conditions in which plastic yielding in the material occurs, at least in some neighborhood of the crack tip.
Plastic effects make stress-strain curves of the material be history dependent. So, after a cyclic variation of the external load, the state of the system may not be identical to that at the beginning of the cycle. Particularly, a finite crack advance in each cycle can occur.
It is important to emphasize that in this case the process does not require the existence of any activation step, allowing a fully deterministic study of the problem once the constitutive elasto-plastic behavior of the material is known.


In general terms, for a mechanical component, a stage of cyclic fatigue propagation can eventually lead to a regime of dynamic propagation, at which abrupt failure occurs. Cyclic fatigue propagation may be the main concern in applications in which mechanical components are exposed to temporally variable loads, or to repetitive thermal cycling. 

We focus on cyclic fatigue propagation in the present paper.
Although the amount of experimental data and the phenomenological treatments of this problem are abundant \citep{suresh.98,bolotin},
the basic studies that are available on cyclic fatigue crack propagation are scarce. The difficulty of the problem is the typical one of fracture mechanics (i.e., the spanning of many orders of magnitude between the process zone at the crack tip to the macroscopic region described by the continuum dynamics) plus the necessity of a detailed description of plasticity
in the material. For instance, in a recent work \citep{farkas.05}, crack advance under cyclic loading
has been obtained in atomistic simulations of a nano-crystalline material. The plasticity in this case is seen to be related to dislocation emission from the crack tip upon loading. Results of this kinds of studies are at present limited by the
availability of computing resources. An alternative approach is to consider the system as described by the equations of
continuum media that include plastic response \citep{finite_elements_2}. 
However, this approach necessarily breaks down sufficiently close to the crack tip, and has to be complemented either by full atomistic simulations 
in a small neighborhood of the tip, or by phenomenological 
prescriptions about the crack tip advance behavior. 

From a conceptual point of view, it is always desirable to have simple models that reproduce the available phenomenology with a minimum of ingredients. 
Having in mind the cyclic loading fatigue crack advance problem, 
the possibility that we have been studying corresponds to spring lattice systems with plasticity in the springs.
In fact, lattice spring models have been a very important benchmark where many predictions of fracture mechanics were tested, and also where effects that go beyond the reach of analytic treatments were obtained \citep{slepyan.81,Marder.95,kessler.99,kessler.01,Guozden.06}. These studies have mainly focused on propagation in lattices with linear, non-linear elastic, or visco-elastic springs. In  all these cases there is a unique relation between a stationary applied strain and the stress produced. To our knowledge there have not been previous attempts to study cyclic fatigue crack growth using lattice spring models with plasticity.

In the present paper we study the simplest case of a lattice of elasto-plastic springs that may cause fatigue crack growth. This is a quasi-one-dimensional model consisting of two identical, parallel elasto-plastic chains, joined by breakable springs, that are laterally pulled apart by the external load, in such a way that a crack can propagate between chains. Formally, this represents a case of ``Mode III" propagation, as we only consider perpendicular  displacements to the chain direction.
In absence of plasticity, this kind of ``one-dimensional" geometry  has been studied in detail in the context of dynamical fracture, and has proved to be useful as a simple benchmark for the more complex behavior that is observed in a more thorough two dimensional implementation \citep{Langer.92,Bouchbinder.08}.
We have also used this model in a previous work \citep{Guozden.06} and obtained results in the dynamical propagation case.
We now concentrate in this quasi one-dimensional model because in addition to the numerical implementation, it allows a detailed, mostly analytical description of the fatigue propagation mechanisms involved, and also because in this case, the germ of experimental features of fatigue crack growth is already observed. We leave for a forthcoming publication the study of a more realistic two dimensional geometry, 
which can be made only through numerical simulation.

In the next Section we describe the quasi-one-dimensional model and the numerical technique in detail. In Section III we present
the main results that we have obtained. They include the very observation of the possibility of fatigue crack advance under cyclic conditions following a Paris-like law, and the observation of the overload retardation effect, in which a cycle using a deformation larger than the average induces an eventual retardation in the advance of the crack. 
Finally in Section IV we discuss the results of our approach in the context of other studies and present the final conclusions.

\section{Details of the model and the numerical technique}

\begin{figure}[H]
\begin{center}
\includegraphics[width=8cm]{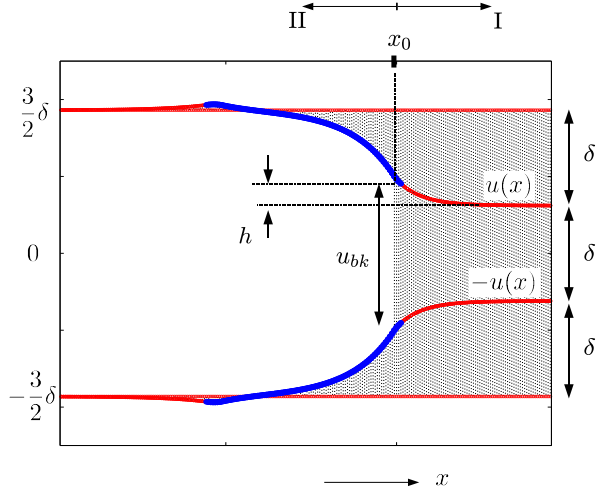}
\end{center}
\caption{A sketch of our idealized system. The crack tip position $x=x_0$ separates regions $\mb{I}$ and $\mb{II}$. Intact inter-chain springs are shadowed. Note that $h$ is given by $h=\frac{\bk-\dd}{2}$. The part of the chains that has accumulated some plastic deformation is shown by a thicker line
(note that the region with plasticity penetrates slightly into region I).}
\label{fig:cadena_fatiga_analitica}
\end{figure}

The system we model is reminiscent of a stripe geometry, under mode III conditions imposed by rigid displacements on the lateral sides, with a crack advancing in the middle of the stripe. In this quasi one-dimensional idealization (see Fig. \ref{fig:cadena_fatiga_analitica}) we consider only two symmetric chains around the middle line of the system: $u(x)$ and $-u(x)$. They are respectively connected through linear springs to the lateral strain gauge located at $\pm 3\delta/2$.  In addition, the two middle chains are connected by breakable springs.  As soon as the length of any of these springs exceeds some threshold value $u_{bk}$, the spring breaks irreversibly. We emphasize that these are the only springs allowed to break in the model. The intact inter-chain springs are shadowed in Fig. \ref{fig:cadena_fatiga_analitica}.

The equations of the model are those of mechanical equilibrium for a given profile of the chain $u(x)$. 
To simplify, we first consider the case in which the continuous profile $u(x)$ is replaced by a discrete set of values $u_j$, where $j$ corresponds to a horizontal coordinate $j\Delta$, $\Delta$ being the discretization parameter. The equilibrium equation for $u_j$ reads

\begin{equation}
 \left (  \frac 32 \delta -u_j\right)\Delta-2\theta  u_j \Delta+F(u_{j+1}-u_j)+F(u_{j-1}-u_j)=0.
\end{equation}
The different terms represent: 

1) the coupling of $u_j$ to the upper border located at $3\delta/2$.

2) The coupling of $u_j$ to the mirror chain at $-u(j)$. In this term the Heaviside function $\theta$ is defined as one, unless $2u_j$ was larger
than $u_{bk}$ at some previous time, in which case it is taken as zero. Note that in the initial condition we take $\theta=0$ for all $j$ lower than some value, to simulate a pre-existent crack. The factor $\Delta$ included in  terms 1) and 2) corresponds to use a spring constant that is  unitary for a unit length of the continuous system.

3,4) the last two terms are the intra-chain forces with the particles at the left and at the right of the site $j$. 
Plasticity is included precisely in these terms, and not in the interchain springs. This choice looks strange at first, since springs
that eventually break are those which experience the largest deformation, and plastic deformation  is expected to be maximum for them. In spite of this, we have found
that the main characteristics of cyclic fatigue can be discussed disregarding plasticity of inter chain springs. In other words, its inclusion does not modify the results in a qualitative way. Since this is the key simplification that allows a detailed analytical treatment, we think that this advantage justifies neglecting plasticity in vertical springs.

We now concentrate in the intra-chain forces. For its description, we split the force $F$ in two parts, an elastic part $f^e$ and an inelastic part $f^p$ where the effects of plasticity will be included. The total force will be $F=(1-P)f^e+Pf^p$ where the parameter $P$ ($0<P<1$)
controls the extent of plasticity of the spring. The elastic part is simply calculated as $f^e=(u_{j\pm 1}-u_{j})/\Delta$. The spring constant is rescaled with the discretization in such a way that it would be one if $\Delta=1$.
To specify the inelastic part,
we introduce a ``rest length" $L^0_{j,j+1}$ for the spring connecting sites $j$ and $j+1$. We calculate the inelastic force as 
$f^p=(u_{j}-u_{j+1}-L^0_{j,j+1})/\Delta$. The total force in the spring is thus given by
\begin{equation}
F=\left[(1-P)(u_{j+1}-u_{j})+P(u_{j}-u_{j+1}-L^0_{j,j+1})\right]/\Delta	
\end{equation}

The force $f^p$ models an ideal plastic behavior. The rest length $L_0$ is initially set to zero to model a virgin sample. Upon small variations of $u_{j+1}-u_{j}$,  $L_0$ is changed, if necessary, to avoid  $f_p$  to go outside some pre-established range $\pm \nl\Delta$, namely, if $|f_p|$ becomes larger than $\nl\Delta$, $L_0$ is adjusted to get 
$|f_p|=\nl\Delta$. We will refer to cases where values of $L_0$  different from zero appear, as cases in which a ``plastic deformation" is present in the system.

\begin{figure}[hbpt!]
a)

\includegraphics[clip,width=3cm]{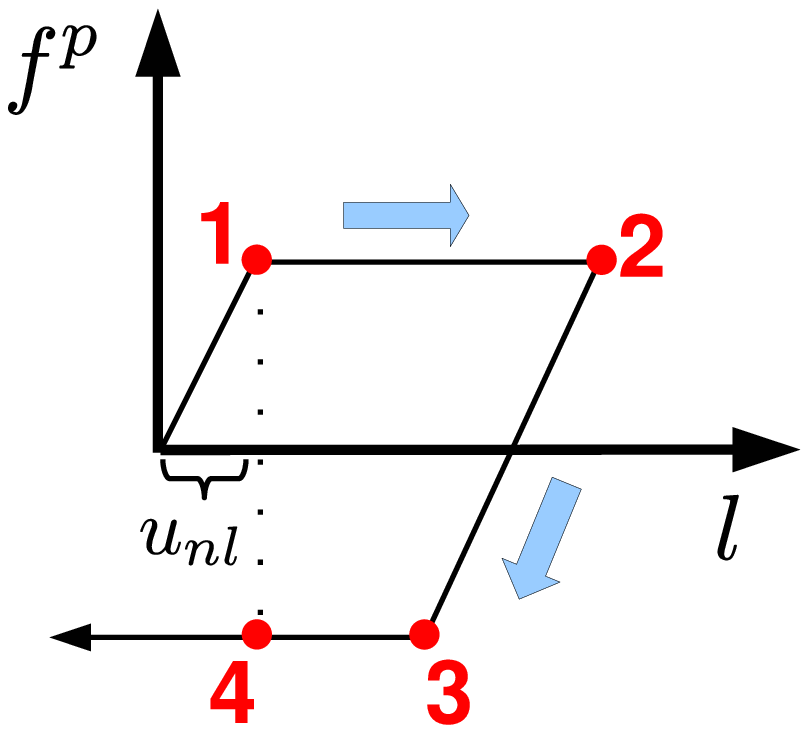}

\vspace{1cm}
b)

\includegraphics[clip,width=3cm]{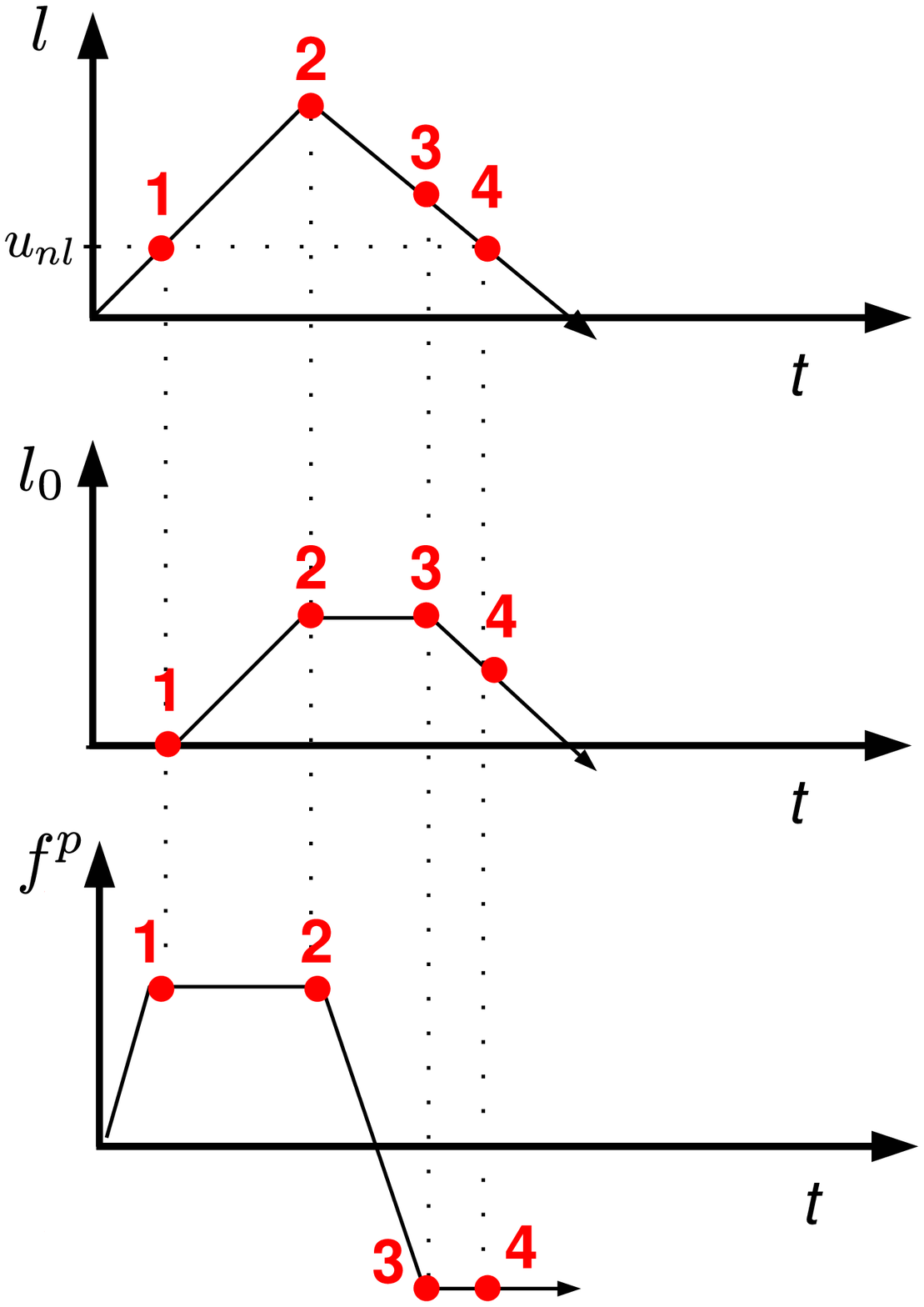}
\vspace{1cm}

c)

\includegraphics[width=4cm]{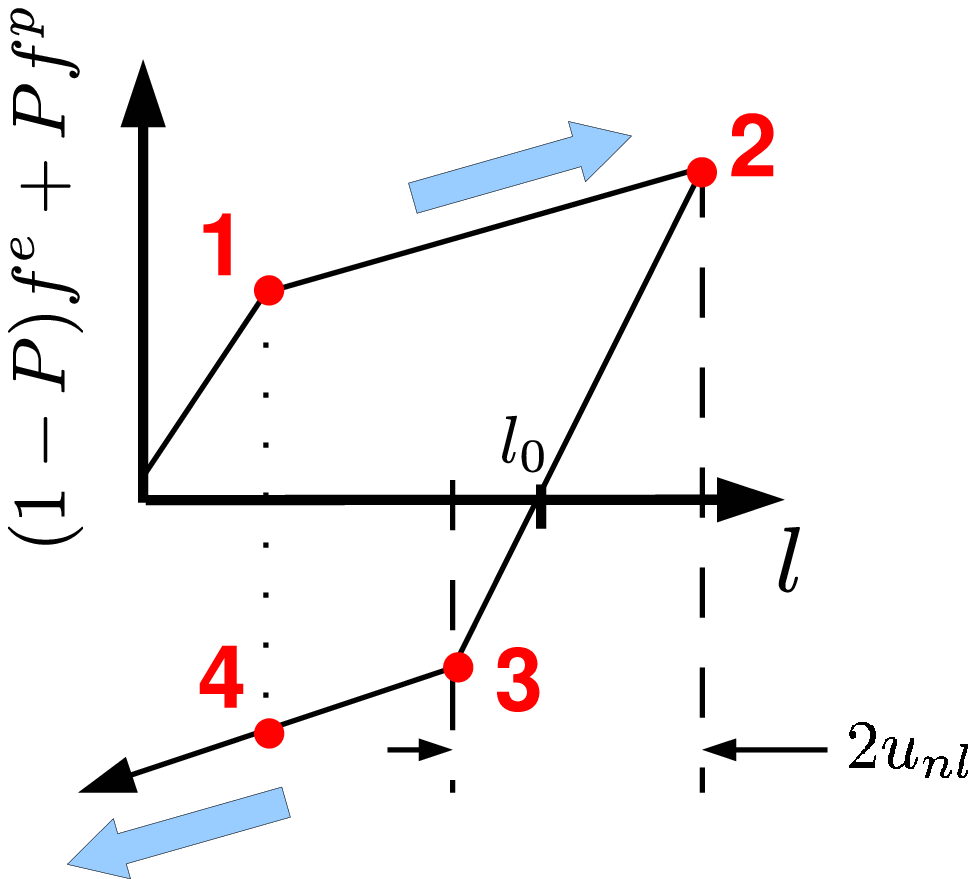}

\caption{Qualitative behavior of the plastic springs. In $a)$ the inelastic force $f_p$
of a spring when $l$ first increases and then decreases is shown.
In $b)$ the corresponding
evolution of length ($l$), rest length ($l_0\equiv L_{0}/\Delta $), and $f_p$,  along this path
is indicated. Note that when the $f_p$ reaches its maximum allowed value, $l_{0}$ starts to change.
In c), the total force, including the elastic part, is indicated (we use $P = 0.5$).}

\label{unl} 

\end{figure}

Through this mechanism, we can see that history dependent forces appear in the system.
In Fig. (\ref{unl}) we show the evolution of $f^p$ and the total force $F$ as the length of the spring $l\equiv u_{j+1}-u_j$ is changed in a prescribed manner.
Note that although the evolution is fully deterministic, the force is not a single valued function of $l$.  For instance, the spring has the same length $l$ at points 1 and 4  but the force it exerts is different.

The numerical procedure we use to find stationary solutions to this problem consists in starting with a  pre-existent crack and no plastic deformation ($L_0\equiv0$) in the system, and increase the value of $\delta$ in small steps. At each step the solution to the equilibrium equations is obtained through a relaxation protocol, and after that the value of $\delta$ is increased again. Once some maximum value of $\delta$ is reached, the same procedure is repeated for decreasing $\delta$. As it was already emphasized, the history dependent stress-strain relation of the springs makes it possible to observe a systematic crack tip advance upon cyclic loading conditions.

The previous description is appropriate for numerical implementation, but for an analytical treatment its is better to return
to the continuous description.
This is achieved by letting the discretization parameter $\Delta$ go to 0. 
The continuous equation for our model is obtained as
\begin{equation}
\frac{d^2u}{dx^2}+\left(\frac{3\delta}{2}-u\right)+ 2\theta u+ P\frac{dl_0}{dx}=0,
\end{equation}
where we have defined $l_0\equiv L_0/\Delta$. In this description, the system is continuous along the chains, although it remains essentially discrete
in the perpendicular direction. This kind of continuous	 limit along a single spatial dimension is well defined \citep{Guozden.06}, contrary to the case of a full continuous limit in which the transition from discrete to continuous is much more subtle. 
Given a distribution of plasticity $l_0(x)$ the previous equation can be solved. However, upon a variation of $\delta$, 
this equation does not stand alone, as it has to be complemented with the prescription that $l_0$ adjust itself 
to satisfy $|du/dx|-l_0(x)<u_{nl}$.

\section{Results}

\subsection{Crack advance upon monotonous load increase}

\begin{figure}[H]

\begin{center}
\includegraphics[width=6cm]{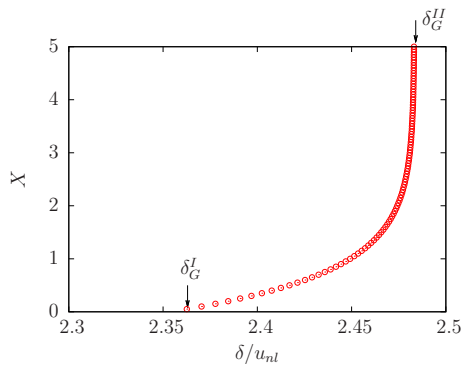}
\end{center}


\caption{Crack tip position $X$ as a function of an increasing value of $\delta$, starting from a virgin configuration (namely, no plastic deformation in the initial state). Parameters are $\bk/\nl=4$, $P=0.5$, and $\Delta=1/20$.
}
\label{avance} 
\end{figure}

In the absence of plasticity ($P=0$), the system is perfectly elastic, and an energy balance analysis can be applied: if the strain $\delta$ is lower
than some critical value $\delta_G$, there is only a static solution in which the crack tip position is stable (assuming crack healing does not occur), as the external load is not able to provide the
necessary energy for crack advance. Instead, for $\delta >\delta_G$, there is in principle enough available energy for the system to be in a run away state, in which the crack moves forward at some finite velocity. This corresponds to a continuous description.  The discreteness of the system introduces the additional ingredient of lattice trapping \citep{Thomson.71,Paskin.81,Kessler.99.lt,Bernstein.03}, and the minimum necessary value of $\delta$  to have a run away solution becomes somewhat larger than the value $\delta_G$ determined by purely energetic arguments.
When plasticity is incorporated into the   springs ($P>0$), a new regime occurs, intervening between the static and dynamic regimes. In fact, at a first critical value of $\delta$ that we call $\dgs$, the crack starts to elongate. However this elongation is not unstable now. Instead, there is a well defined value of the crack tip advance $X$ as a function of $\delta$. The crack only runs away when a second critical value $\dge$ is exceeded. The numerical results that show this behavior are presented in Fig. \ref{avance}. For any value of $\delta$ between $\dgs$ and $\dge$ the crack tip position is stable.
%

\begin{figure}[H]
%
%
%
%
%
%
%

\includegraphics[width=7cm]{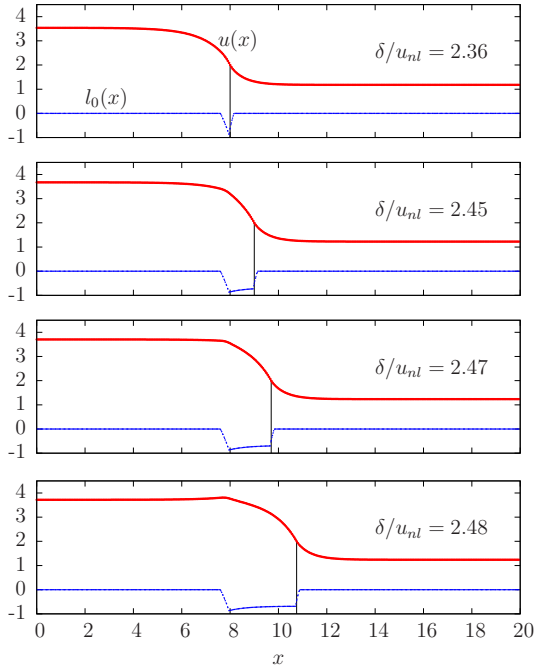}
\caption{Upper curves: chain profiles as $\dd$ increases. Lower curves: corresponding plastic deformation $l_0$ of the springs. The thin vertical lines indicate the crack tip position. Note that for large values of crack tip advance, the value of $l_0$ behind the crack tip reaches an asymptotic value 
(parameters as in Fig. \ref{avance}).}
\label{fig:esquemas_cadena_fatiga}
\end{figure}

\begin{figure}[H]
\begin{center}
\includegraphics[width=5cm]{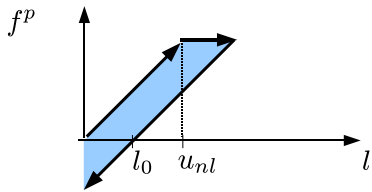}
\end{center}
\caption{Inelastic force on the springs, as the crack passes by. The shadowed area corresponds to the energy used in the plastic deformation, that
amounts to $P(l_0 u_{nl}+l_0^2/2)$.}
\label{fig:ener_plas}
\end{figure}

The actual configurations of the system at increasing values of $\delta$ are presented in Fig. \ref{fig:esquemas_cadena_fatiga}. There we plot the chain profile $u(x)$, and also the corresponding values of the rest length $l_0(x)$ of the springs forming the chain.
Upon crack tip advance, the portion of the chain that is left behind remains with a finite plastic deformation (i.e. $l_0\ne 0$ behind the crack tip). 
This is of the utmost importance as it implies the necessity of some energy expenditure. In fact, Fig. \ref{fig:ener_plas} shows the typical variation of the plastic force on each spring after the crack tip passed. The area of this curve is the energy dissipated in the system due to plastic deformation. This is the reason why the crack does not immediately destabilize when $\delta$ exceeds the first threshold $\dgs$. 

We proceed now to a detailed analytical study that gives insight into the process of quasistatic crack advance upon load increase.
In order to obtain analytical 
expressions, we have to consider the continuous limit, namely, the case $\Delta \to 0$.
First of all, we calculate  $\delta_G^I$ and $\delta_G^{II}$ in this limit. 

The calculation of $\delta_G^I$ can be made as if springs were non-linear elastic, since plasticity is not involved. In fact, upon increasing $\delta$, as long as no spring is broken (as it happens below $\delta_G^I$), all spring lengths also increase.
This means that springs explore only the ascending part in Fig. \ref{unl}$a$, which corresponds to a non-linear elastic behavior. In these conditions, the threshold can be determined by an energy balance argument, equating the available energy ahead of the crack with the energy that is necessary to break a spring. This gives the expression for $\delta_G^I$ in our geometry
\begin{equation}
3\frac{(\dgs)^2}{2}=\frac{\bk^2}{2},
\end{equation}
from which
\begin{equation}
\delta_G^I=\frac{\bk}{\sqrt 3}.
\end{equation}
Note that this value does not depend on the  parameter $P$.

As soon as the crack tip moves on, some springs reduce their length, exploring the descending part of the curve in Fig. \ref{unl}a, and plasticity becomes involved. 
The calculation of $\delta_G^{II}$ can still be made  through an energy balance argument, if we
take into account the additional energy spent in the process. For each individual spring within the chain, its evolution during the whole process  implies an initial stretching, and then a reversion to a state of zero length, as indicated in  Fig. \ref{fig:ener_plas}. The energy $E_0$ spent in this process is proportional to the shadowed area in the figure, and its value is $E_0=P(l_0 u_{nl}+l_0^2/2)$. 
Knowing the value of $l_0$, $\dge$ can be calculated from the energy balance equation
\begin{equation}
3\frac{(\dge)^2}{2}=\frac{\bk^2}{2}+2E_0(l_0)
\label{balance}
\end{equation}
(the factor 2 in the last term comes from the fact that energy is spent in the plasticity of springs in the two mirror chains).

The value of $l_0$ can be calculated through the following argument: as $\dd\rw\dge$ 
every spring behind the crack tip gets a plastic deformation $l_0$ \footnote{This is true if  $\frac{\bk}{\nl}<2+\sqrt{12-9P}$. We restrict to this case throughout this work.}. This means that this part of the chain is effectively linear elastic. Its profile is given by 
\begin{equation}
u(x)=\frac{3\delta}{2}-\left(\frac{3\delta}{2}-\frac{u_{bk}}{2}\right)e^x
\end{equation}
where the slope of the chain at the crack tip (located at $x_0=0$) is $\partial u/\partial x=(\bk-3\dd)/2$. Then, the plastic deformation at the crack tip is, in norm,
\begin{equation}
|l_0|= \frac{3\dd-\bk}{2}-\nl.
\end{equation}
Inserting this result into Eq. (\ref{balance}) for $\delta\to \dge$ we get
\begin{equation}
{\dge}^2=\frac{\bk^2}{3}+\frac {2P}3\left[ \left(   \frac{3\dge-\bk}{2}\right)^2-\nl^2\right].
\label{deltag2}
\end{equation}
It can thus be seen that $\dge$ coincides with $\dgs$ for $P=0$, and 
becomes progressively larger when $P$ increases, allowing a finite range $\dge-\dgs$ of stable crack elongation.
For the particular value $P=0.5$ used in previous figures the value of $\dge$ is
\begin{equation}
{\dge}=-\bk+\frac{2}{\sqrt{3}}\sqrt{2\bk^2-\nl^2}
\label{deltag3}
\end{equation}

Now we derive an equation from which the full curve $X(\delta)$ can be obtained.
We will refer to the definitions in Fig. \ref{fig:cadena_fatiga_analitica}. In particular we consider
region $\mb{I}$ as that ahead of the crack tip ($x>x_0$), and region $\mb{II}$ as that behind the crack tip ($x<x_0$). 

Region $\mb{I}$: Here the horizontal springs deform monotonically, so again plasticity does not take part. We only have to consider that springs behave as piecewise linear, with two different spring constants, for elongation below and above $u_{nl}$.
This offers the possibility to obtain a  solution in this region, and to evaluate the slope of the chain at the crack tip.
\footnote{Here we assume that the chain actually explores the nonlinear regime. This happens when $|\frac{\bk-\dd}{2}\sqrt{3}|>\nl$}
After some algebra we obtain
\begin{equation}
\left. \frac{du_{I}}{dx} \right|_{x=x_0}=\frac{-\sqrt 3}{\sqrt{1-P}} \sqrt{\left(\frac{\bk-\dd}{2} \right)^2- \frac{P}{3}\nl^2 }
\equiv R(\delta),
\label{eq:derivada_zona1}
\end{equation}
where we have renamed the expression to $R(\delta)$ for future use.

In region $\mb{II}$ we proceed as following: 
As we indicated previously in the introduction, the existence of a non-zero $l_0(x)$ function can be incorporated into the equilibrium equation of the system as and additional force $Pdl_0/dx$.
This means that the equation describing the deformation $u(x)$ in region $\mb{II}$ is
\begin{equation}
u''(x)+\left(\frac{3\dd}{2}-u(x)\right)+P\frac {dl_0}{dx},
\end{equation}
with boundary conditions
\begin{equation}
u(-\infty)=\frac{3\dd}{2},~~~~~~u(x_0)=\frac{\bk}{2}.
\end{equation}

Solving formally this equation we find the derivative of the chain at the crack tip in terms of $l_0(x)$. We obtain
\begin{equation}
\left.\frac{du_{II}}{dx}\right|_{x=x_0}=P \int_{-\infty}^{x_0} \frac {dl_0}{dy}e^{-(y-x_0)}dy-\frac{3\dd-\bk}{2}.
\label{eq:derivada_zona2}
\end{equation}
It is interesting to emphasize the meaning of this equation, stating that any non-homogeneous $l_0(x)$ behind the crack tip contributes to the slope of the chain at the crack tip, and its effect decays exponentially with distance.
Equating the results for regions $\mb{I}$ (Eq. \ref{eq:derivada_zona1}) and $\mb{II}$ (Eq. \ref{eq:derivada_zona2})
we obtain 
\begin{equation}
P\int_{-\infty}^{x_0} \frac{dl_0}{dy}e^{-(x_0-y)}dy-\frac{3\dd-\bk}{2}=R(\dd).
\end{equation}
\vspace{.5cm}

To solve this integral equation, we multiply each of the terms by $e^{x_0}$ and then take a derivative upon a variation in the crack tip position $x_0$.
In doing this, we also use the fact that upon an advance of the crack tip, the values of $l_0$ are not modified behind the crack tip.
\footnote{Same as  $^{1}$}
The result is
\begin{equation}
P\frac{dl_0}{dx_0} e^{x_0} -\frac32 \frac{d\dd}{dx_0} e^{x_0}-\frac{3\dd-\bk}{2}e^{x_0}=\left[\frac{dR(\dd)}{dx_0}+R(\dd)\right]e^{x_0}
\label{eq:qersimplif}
\end{equation}

The value of $l_0$ at the crack tip in the first term $l_0(x_0)$ can be written as $l_0(x_0)= \left.du/dx\right|_{x=x_0}-u_{nl}=R(\delta)-u_{nl}$
and after replacing in equation (\ref{eq:qersimplif}) we obtain
\begin{equation}
P\frac{dR(\dd)}{dx_0}  -\frac32 \frac{d\dd}{dx_0} -\frac{3\dd-\bk}{2}=\frac{dR(\dd)}{dx_0}+R(\dd)
\label{eq:en_terminos_dd}
\end{equation}
Changing from the strain variable $\delta$ to $h$, as defined in Fig. (\ref{fig:cadena_fatiga_analitica}), we arrive to
\begin{equation}
P\frac{dR(h)}{dx_0}  +3\frac{dh}{dx_0} +3h-\bk=\frac{dR(h)}{dx_0}+R(h),
\label{eq:R1}
\end{equation}
where $R(h)$ is now
\begin{equation}
R(h)=-\frac{\sqrt{3h^2- P\nl^2 }}{\sqrt{1-P}}.
\end{equation}
In addition, using Eq. (\ref{eq:derivada_zona1}), the derivative of $R$ with respect to $x$ is
\begin{equation}
\frac{dR(h)}{dx}=\frac{3}{1-P}\frac{h \frac{dh}{dx}}{R(h)},
\end{equation}
and combining with Eq. (\ref{eq:R1}) we obtain
\begin{equation}
\frac{dh}{dx_0}=\frac{\frac{R(h)}{h}-3}{3-3\frac{h}{R(h)}}h+\frac{\bk}{3-3\frac{h}{R(h)}}.
\label{diferencial_h_vs_xo}
\end{equation}

We have finally obtained a differential equation linking strain (contained in $h=(u_{bk}-\delta)/2$) and  the crack tip position $x_0$. 
The equation is nonlinear, and we cannot give an analytical solution in a general case, but many general features can be worked out.
First of all it can be checked that the right hand side of Eq. (\ref{diferencial_h_vs_xo}) vanishes linearly at
$\delta=\delta_G^{II}$. This indicates that the crack tip advance $X(\delta)$ has a logarithmic divergence at $\delta_G^{II}$.
Second, we note that Eq. (\ref{diferencial_h_vs_xo}) will determine the crack tip position as a function of $\delta$ up to an additive
constant, i.e, the solution will be of the form 

\begin{equation}
x_0=F(\delta)+A,
\label{a}
\end{equation}
with $F(\delta)$ a well defined function diverging logarithmically at $\delta_G^{II}$, and $A$ an arbitrary constant. The value of $A$ can be set by using the condition that $X$ becomes different from zero exactly at $\delta_G^I$ for a virgin sample. This defines the solution unambiguously. 

Expanding $R(h)$ (see equation (\ref{eq:derivada_zona1})) in powers of $h$, 
we obtain to linear order
\begin{equation}
R(h)\simeq -\sqrt  {\frac{3}{1-P} }~~h,
\label{rlineal}
\end{equation}
which becomes exact as ($P\to0$).

Using this approximation for $R(h)$ in (\ref{diferencial_h_vs_xo}) leads to the following linear first order differential equation
\begin{equation}
\frac{dh}{dx_0}=-\left(\frac{\sqrt 3 +\sqrt{\frac{1}{1-P}} } {\sqrt 3 + \sqrt{1-P}}\right)h+\frac{\bk}{3+\sqrt3\sqrt{1-P}  }.
\label{eq_linear}
\end{equation}
In this approximate case, the solution is fully described by a logarithmic dependence of $x_0$ as a function of $\delta$, rising linearly from zero at $\delta_G^I$, and diverging at $\delta_G^{II}$.

\begin{figure}[hbtp!]
\begin{center}
\includegraphics[width=7cm]{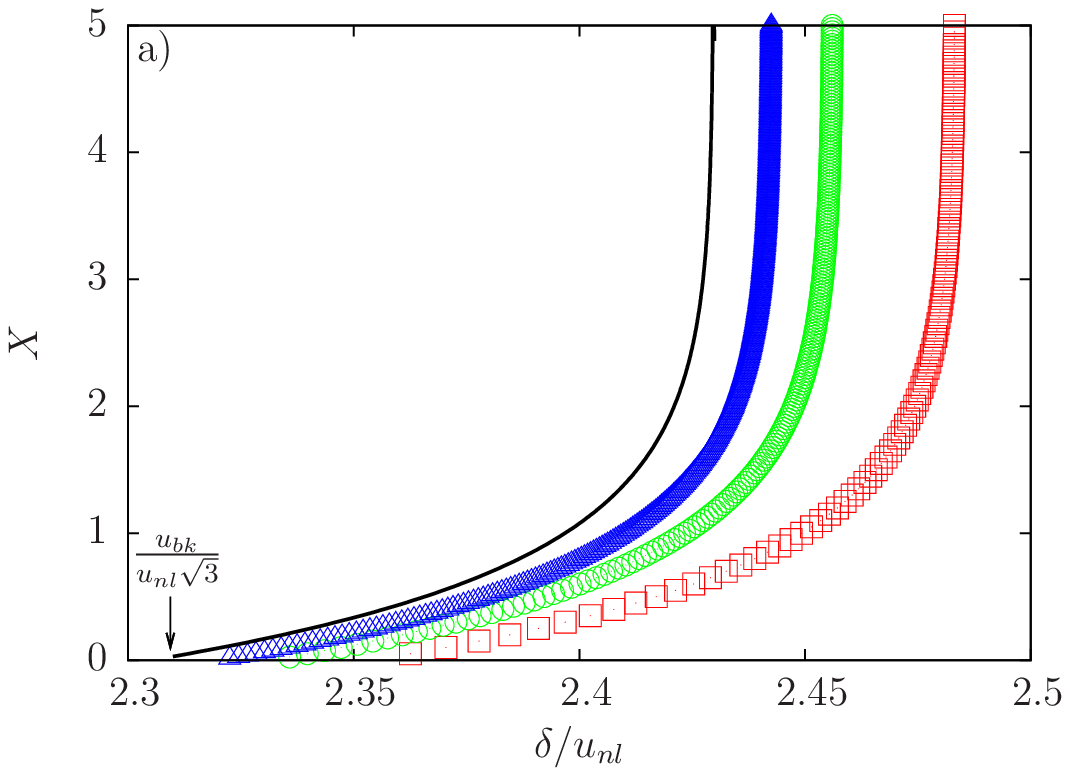}
\end{center}

\begin{center}
\includegraphics[width=7cm]{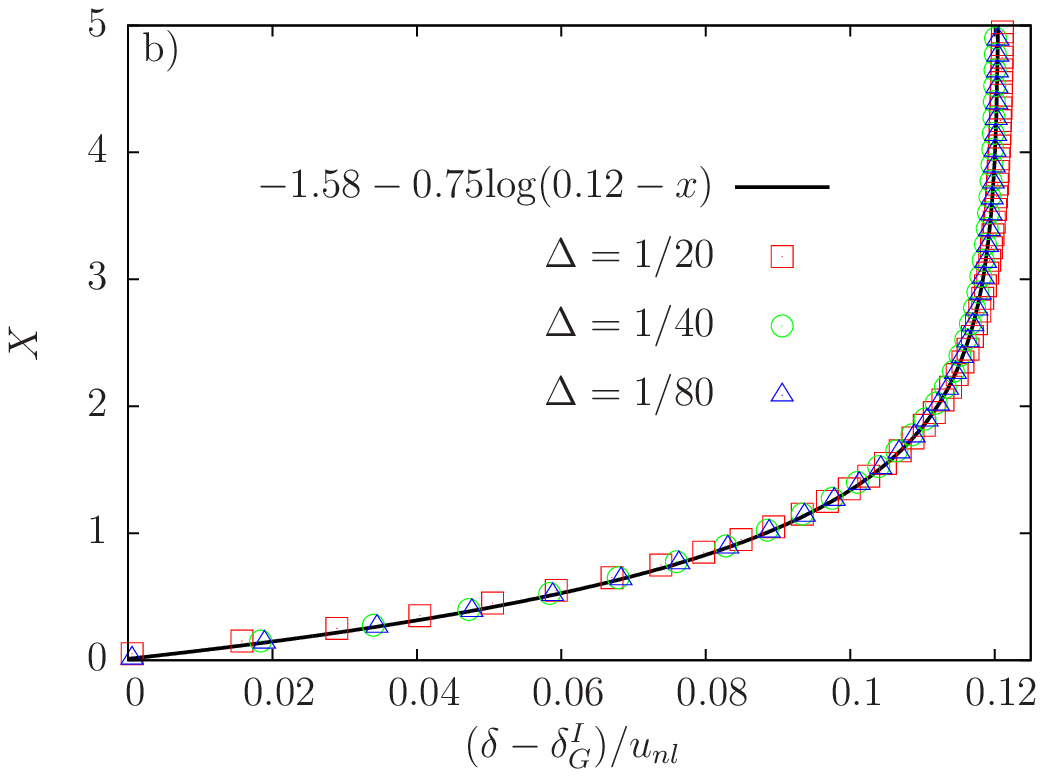}
\end{center}

\caption{
(a) Analytical results for the crack tip advance $X$ according to the linear approximation given in Eq. (\ref{eq_linear}) and results of numerical simulations at different values of $\Delta$.
(b) The same results shifted according to the values of $\delta_G^I$ ($\bk/\nl=4$, $P=0.5$).
}
\label{fig:avance_epsilon}
\end{figure}

To check this behavior, a comparison of the analytical expressions obtained with results of numerical simulations is presented in Fig. \ref{fig:avance_epsilon}. The numerical simulations show an important dependence on the parameter $\Delta$, which can be considered in terms of a lattice trapping effect \citep{Thomson.71,Paskin.81,Kessler.99.lt,Bernstein.03}. The continuous limit is obtained by letting $\Delta\to 0$. We see that when this is taken into account the agreement between the analytical and numerical results is very good.

\begin{figure}[H]
\includegraphics[width=8cm]{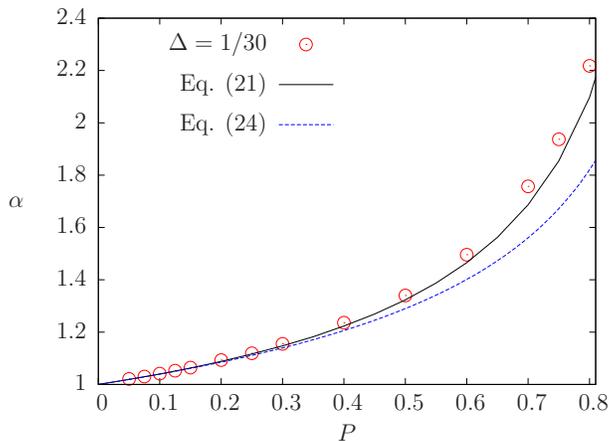}
\caption{Results for the pre-factor $\alpha$ of the logarithmic divergence of $X(\delta)$ at $\delta_G^{II}$. Circles are obtained fitting exponential functions to our numerical results with $\bk/\nl=4$, and $P=0.5$. 
Continuous line is obtained
from the full Eq.  (\ref{diferencial_h_vs_xo}), and dotted line is the result of the linear approximation (Eq. \ref{eq_linear}).
}
\label{fig:exponente_vs_P}
\end{figure}

In cases in which we cannot use the linear approximation (\ref{rlineal}) for $R(h)$, Eq. (\ref{diferencial_h_vs_xo})  has to be treated numerically. 
For instance, the slope of the right hand side of Eq. (\ref{diferencial_h_vs_xo}) at zero crossing determines the pre-factor of the logarithmic divergence at $\delta_G^{II}$.
The comparison of this pre-factor with results obtained from full numerical simulations of the model are contained in Fig. (\ref{fig:exponente_vs_P}).
We observe an almost perfect agreement when using Eq. (\ref{diferencial_h_vs_xo}), and deviations if using the linear approximation Eq. (\ref{eq_linear}).

\subsection{Cyclic crack growth and Paris curves}

Crack tip advance upon monotonous increase of the applied load, analysed in the previous section, is the starting point for the study of the more interesting regime of cyclic loading conditions. Particularly we consider the case in which, after the initial application of some  $\delta_{max}$, the stretching is reduced to some $\delta_{min}$, and then it is successively cycled between these two values.
We  expect a finite crack propagation in each cycle of this process. For this to happen, a crucial condition must be fulfilled: 
in the unloading part of the cycle, some of the plastic deformation must be reverted. Otherwise, if downloading is totally elastic, the next loading half cycle will be elastic too, and crack tip advance will not occur.
Reverted plasticity allows a sustained advance in successive cycles. In fact, imagine the hypothetical situation in which all plastic deformation
induced during the first loading is reverted upon unloading, namely the $l_0$ of all springs reset to zero value. This would mean that after a complete cycle, the system is back in the original ``virgin" state, with the only difference that crack tip has moved forward, and the situation will be repeated cycle after cycle.

\begin{figure}[hbtp!]
a)
\includegraphics[width=6cm]{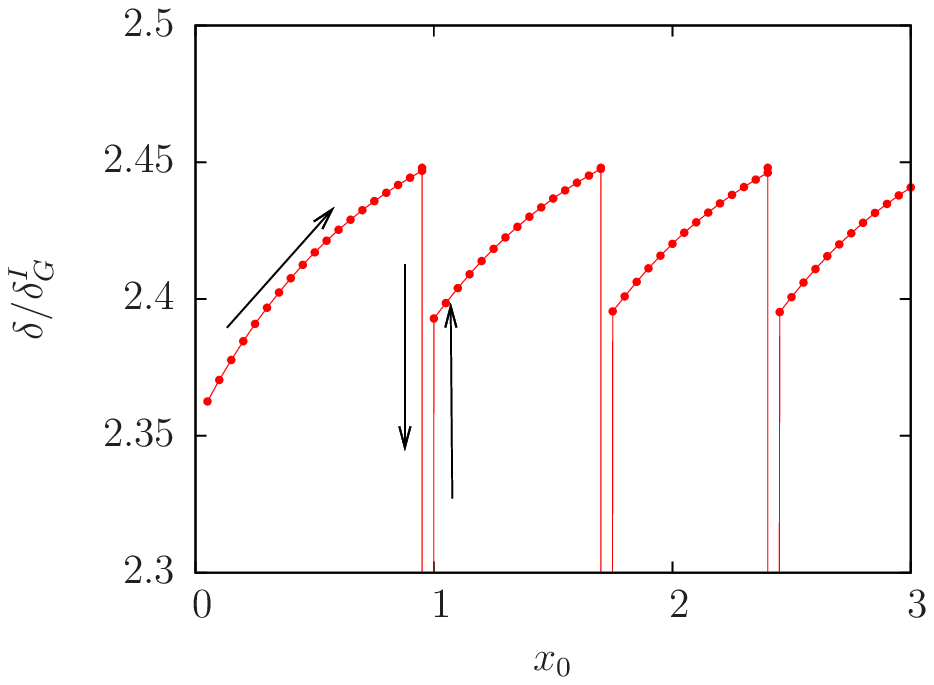}

b)
\includegraphics[width=8cm]{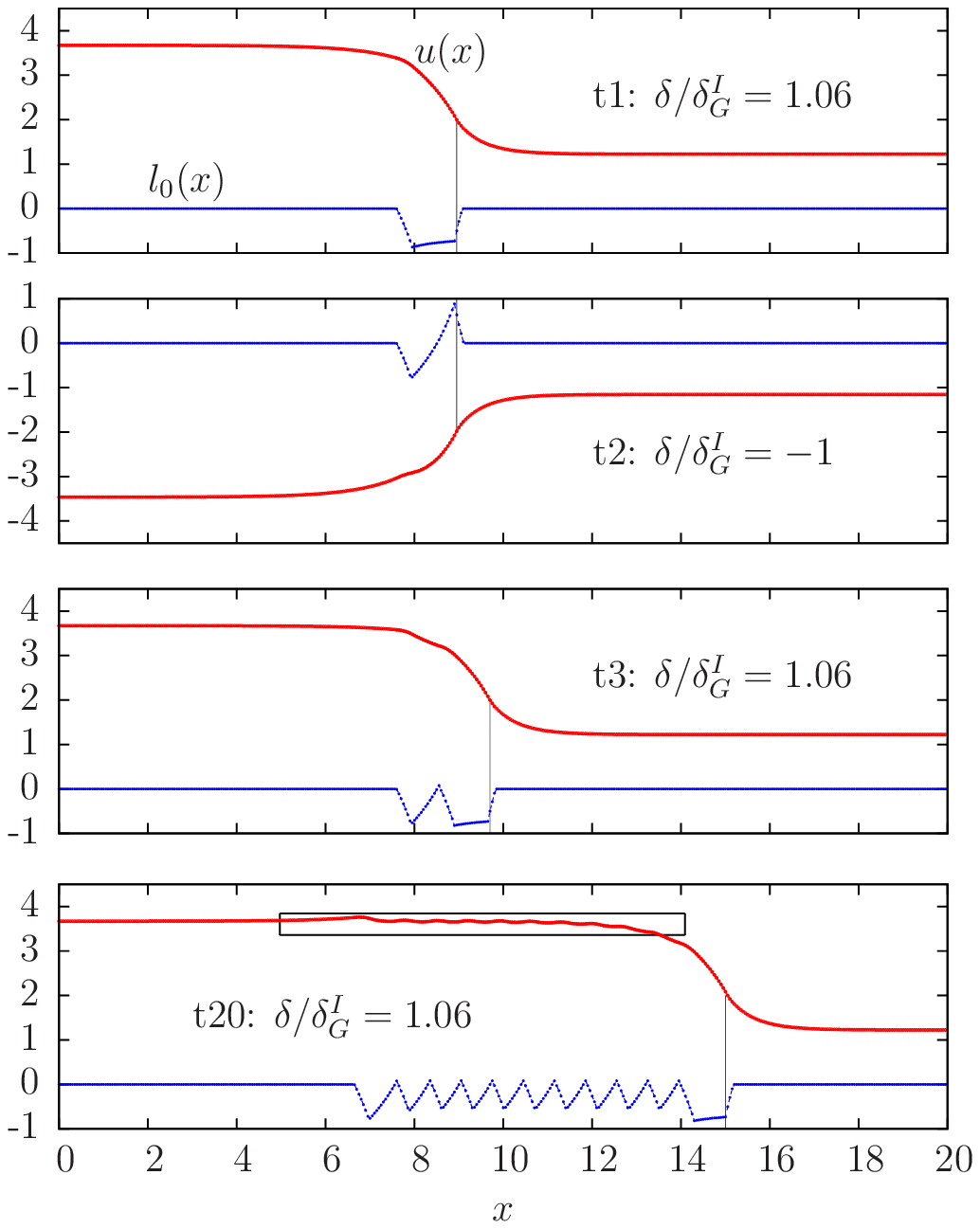}

c)
\includegraphics[width=6cm]{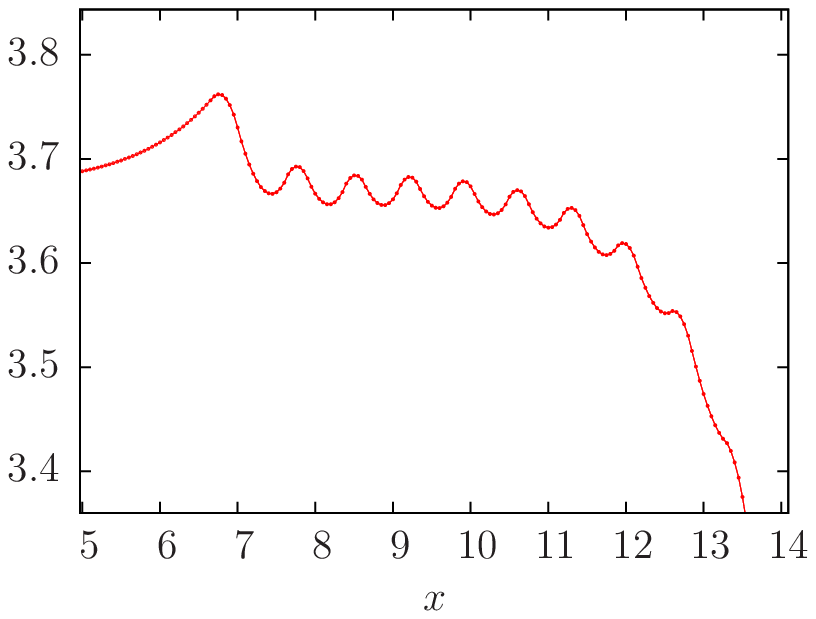}

\caption{$(a)$  Temporal evolution of crack tip position upon cyclic change of load, starting with a virgin sample. We see that after a few cycles, the advance per cycle stabilizes to a finite quantity. (b) Snapshots of the system during the process. Panels are snapshots at the first $\delta_{max}$, the first $\delta_{min}$, the second $\delta_{max}$, and at $\delta_{max}$ after many cycles.
The typical serrations of cyclic fatigue advance are observed in the plastic deformation profile behind the crack tip and in the geometrical profile of the chain itself, as panel $(c)$ shows.
Parameters used are  $P=0.5$, $\Delta=1/20$, $\bk/\nl=4$, $\delta_{min}=-\delta_G^I$ and $\delta_{max}=1.06 \delta_G^I$.}
\label{fig:ciclos_vs_tiempo}
\end{figure}

The numerical results within our model fit in between these two limits.
They are presented in Fig. 
\ref{fig:ciclos_vs_tiempo}. 
Starting from a virgin sample, we first reach some value $\delta_{max}$ in between $\dge$ and $\dgs$. This generates a first crack elongation and a plasticity wake that was discussed in detail in the previous section. Then $\delta$ 
is reduced to $\delta_{min}$. If during this reduction of $\delta$ some part of the plasticity in the system is reverted, the system can elongate the crack further in a second increase of $\delta$. Whether this plasticity reversion occurs or not, depends a great deal on the value of $\delta_{min}$. If this value is too high it does not occur. In the case of the present model, we have seen that we have to take 
$\delta_{min}$ negative in order to have reverted plasticity. For this reason in Fig. \ref{fig:ciclos_vs_tiempo} we use $\delta_{min}=-\delta_G^I$.

A brief digression is convenient at this point. The use of $\delta$ values of both signs is perfectly allowed in our mode III configuration. However, having in mind
an experimental situation in mode I configuration, where $\delta$ should be strictly possitive, we may worry about the neccessity to include values of $\delta$ of alternating signs in the present case in order to observe fatigue crack propagation. 
In this respect, we mention that the need of a change of sign of $\delta$ is an artifact of the quasi
one dimensional system. In two dimensional (i.e., many chains) mode III simulations we have observed reverted plasticity and cyclic fatigue crack advance for strictly positive values of $\delta_{min}$.

In Fig. \ref{fig:ciclos_vs_tiempo} we see that part of the plasticity is reverted during $\delta$ reduction. We stress, however, that for the current parameters there is no crack advance during the stress reduction half period. Then upon a new increase of $\delta$, 
crack elongation resumes at some $\delta<\delta_{max}$. The form of the $x_0 (\delta)$ crack tip advance curve in successive cycles can be described by assuming that previous cycles only influence the advance in the present cycle by shifting the value of $\delta$ at which elongation resumes. This is understood in terms of the analysis of the previous section: Eq. (\ref{diferencial_h_vs_xo}) is still valid to analyze the crack tip advance in the presence of plasticity behind the crack tip. The only difference is that now the determination of the constant $A$ in Eq. (\ref{a}) cannot be done {\em a priori}, and the plasticity in all the region behind the crack tip has to be taken into account. However, once the crack tip starts to elongate, its evolution is dictated by Eq. (\ref{diferencial_h_vs_xo}). We will not attempt here to explicitly calculate the value of $\delta$ at which elongation starts in each cycle, in terms of the plasticity distribution behind the crack tip. Instead, we restrict to a rather qualitative description.

After a few cycles, the elongation of the crack converges to a fixed amount per cycle. 
The asymptotic value of advance per cycle is seen to be lower than the advance in the first cycle, when there is no plasticity in the system. This 
is consistent with the fact that plasticity is only partially reverted upon unloading.
Serrations in the profile of plastic deformation are clearly visible in $l_0(x)$ (Fig. \ref{fig:ciclos_vs_tiempo}). They manifest also in the chain profile itself, as the last panel of the figure allows to observe.
This is reminiscent of the same texture observed in crack surfaces generated by cyclic fatigue (see for instance \citet{suresh.98}, Fig. 10.4, and \citet{Hertzberg.96}, Fig. 13.11).

\begin{figure}[hbpt!]
%
%
\begin{center}
a)
\includegraphics[width=7cm]{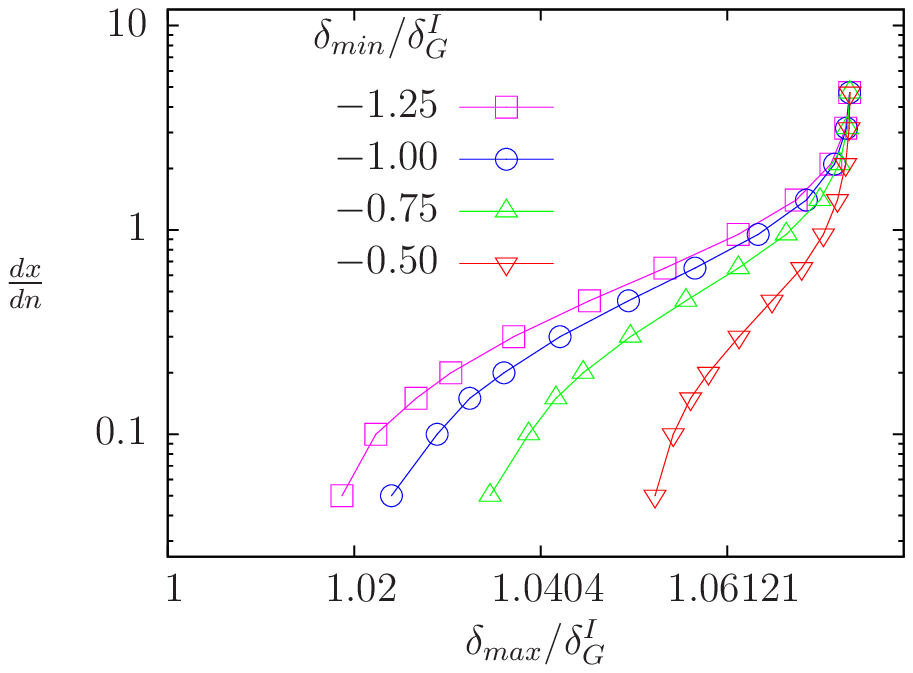}

b)
\includegraphics[width=7cm]{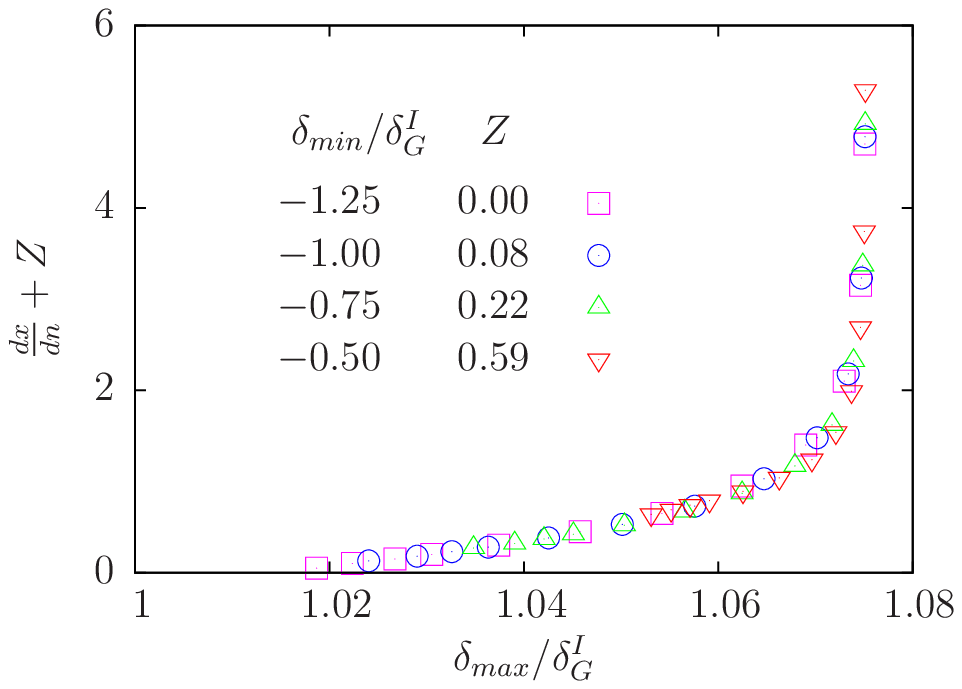}
\end{center}
\caption{
(a)Stationary crack tip advance per cycle $dx/dn$ as a function of $\delta_{max}$, for different values of $\delta_{min}$,
as indicated. The curves show an overall form qualitatively compatible with a Paris law.
In (b), we see that results for different values of $\delta_{min}$ can be absorbed by a vertical shift of the different curves (the vertical shift applied
is indicated).
}
\label{fig:paris_cadenas}
\end{figure}

By systematically running the model at different values of $\dd_{min}$ and $\dd_{max}$, Paris-like curves can be constructed,
displaying the stationary advance-per-cycle as a function of load amplitude.
In Fig. (\ref{fig:paris_cadenas}) we present these results as a function of  $\dd_{max}$, for four different minimum load values $\dd_{min}$, in the usual Paris form, and also in a more appropriate display to our analysis.
We observe that the full curves can be reasonably described by a logarithmic dependence with $\delta_G^{II}-\delta_{max}$, with only a rigid vertical shift to account for the different values of $\delta_{min}$, namely, the advance per cycle $d x/dn$ has the qualitative form

\begin{equation}
\frac{dx}{dn}\simeq \alpha\log(\delta_G^{II}-\delta_{max}) +f(\delta_{min})
\label{quali}
\end{equation}
This expression cannot be put in the form of a dependence on a single combination between $\delta_{max}$ and $\delta_{min}$, as traditional interpretations of the Paris law require. It is however compatible with two-parameter interpretations of the Paris law, as suggested for instance by \citet{Sadananda.94,Sadananda.99,Sadananda.04,Vasudevan.07}.

Note that there is a well defined value of $\delta_{max}$ (depending on $\delta_{min}$) below which there is no systematic crack advance.
This indicates the existence of an endurance limit for our model below which fatigue crack propagation does not occur. The rate of crack advance increases with $\delta_{max}$, and becomes very large approaching $\delta_G^{II}$. In the middle, there is a transition between the two limiting cases. However, this intermediate regime (which is the typical Paris regime) is very narrow compared to most experimental situations (see for instance \citet{Broek.82}, Fig 10.1). We associate this fact to the existence of a single chain in our model.
If the system was ``two dimensional" (i.e., with many chains), plastic deformation would appear in all the chains, and the load region in which fatigue crack advance occurs would be largely increased. We have obtained this enhancement effect in two dimensional simulations, and plan to report on it elsewhere.
For our present simulations, it is remarkable that such a simple model displays the same qualitative behavior observed in a large class of real materials.

\subsection{Overload retardation}

A characteristic effect observed in cyclic crack growth is overload retardation (\citet{suresh.98}, pp520-526, \citet{bolotin}pp.257-263, \citet{Sadananda.99}). Consider a crack subjected to cyclic loading, advancing some finite distance per cycle. If the maximum load in one particular cycle is increased, the crack will advance a larger distance in that cycle. But also a relatively bigger  plastic deformation will be produced, that will shield the crack tip and reduce the advance in the following cycles so that eventually, when  the system reaches steady state advance again, the crack tip position may be lagged with respect to the case in which overload was not present. In some cases, the crack may even completely arrest after the overload. This somewhat counter intuitive phenomenon is very well known experimentally, and is a stringent constraint for any model that is supposed to describe fatigue crack growth. Models
that have been proposed to describe this behavior may be classified into two main categories.
In crack tip plasticity models \citep{Wheeler.72, Willenborg.71} it is assumed 
that crack growth retardation occurs due to the large plastic zone developed during
overloading. The residual compressive stresses formed in this zone will reduce the 
magnitude of the tensile stresses during the next fatigue cycle and tend to delay crack   
growth.
In crack closure models \citep{Elber.71} it is argued that as a result of the tensile plastic 
deformation left in the wake of a fatigue
crack, a partial closure of the crack faces 
occurs during part of a fatigue load cycle.

\begin{figure}[hbpt!]
a)

\includegraphics[width=7cm]{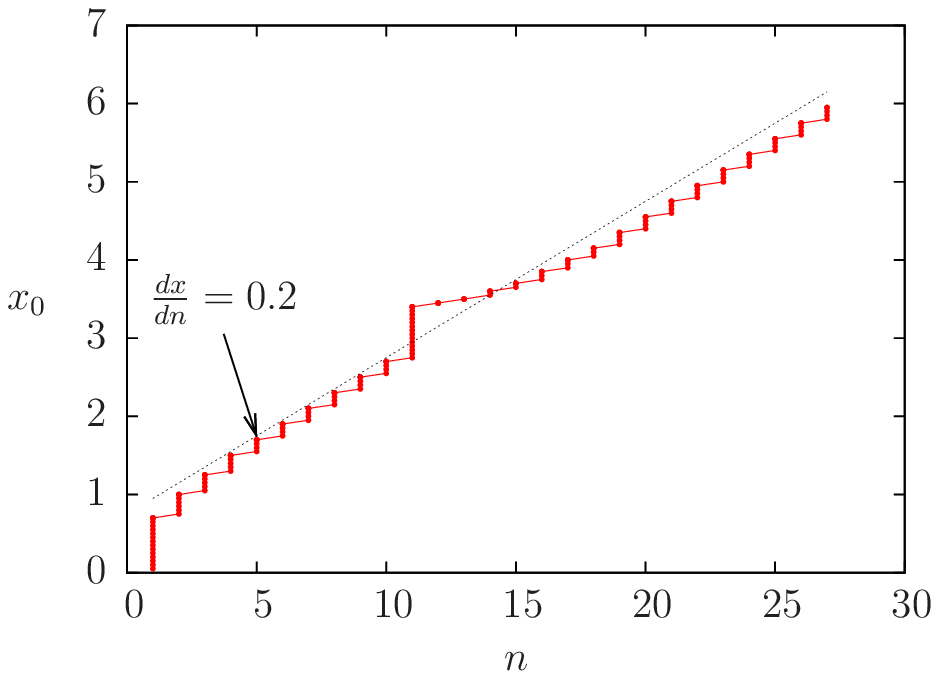}

b)

\includegraphics[width=8cm]{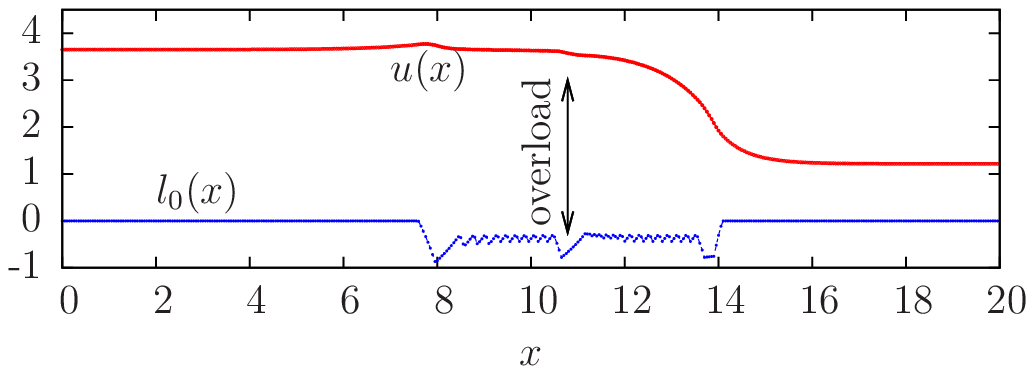}

%

\caption{Overload effect: $a)$ crack tip position versus number of cycles. The crack is loaded cyclically between $\delta_{max}=1.054\delta_G^I$
and $\delta_{min}=-0.6\delta_G^I$. At $n=11$ an overload $\delta_{overload}=1.063\delta_G^I$ is applied. The effect after many cycles amounts to a net retardation of two lattice units, in the present case.
$b)$ Profile of the system $u(x)$ and the springs rest length $l_0(x)$ in the final configuration (simulations with $\Delta=1/20$). Compare panels a) and b) with Figures 10.10, and 2.29 in \citet{Broek.82}.
}
\label{fig:overload_xn}
\label{fig:esquema_overload}
\end{figure}

We have been able to reproduce the overload retardation effect using our model (Fig. (\ref{fig:overload_xn})). The system is initially cycled between $\delta_{max}=1.054\delta_G^I$ and $\delta_{min}=-0.6 \delta_G^I$. After about $n\approx 10$ cycles the growth rate stabilizes at $dx/dn=0.2$ which corresponds to four lattice units per cycle, taking into account that $\Delta=1/20$ in this simulation. In cycle $n=11$ 
the overloading is applied, through 
an increase in $\dd_{max}$ of about 1\%, and the crack is seen to advance more than 3 times the previous value. In the following cycles the crack advance is seen 
to be reduced drastically, and the whole effect once the original rate is recovered results in a net retardation of 2 lattice units.
The reason for this reduction in crack tip advance is the larger plastic deformation induced during the overload cycle.
Fig. (\ref{fig:overload_xn})b shows this enhanced plastic deformation in $l_0(x)$ at the overload cycle. The additional plastic deformation has
a stabilizing effect on the crack tip, so the application of $\delta_{max}$ after the overload is much less efficient in generating crack tip advance. Eventually, when the crack tip has moved far away of the enhanced plasticity zone, the crack tip elongation rate returns to its equilibrium value. 

\begin{figure}[hbpt!]
\includegraphics[width=8cm]{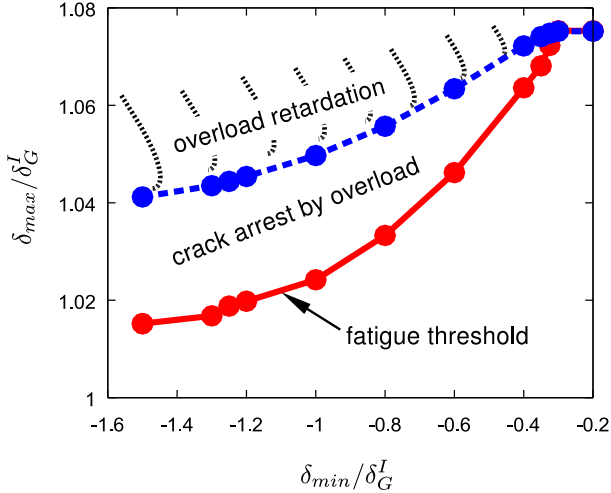}
\caption{The plane $\delta_{max}$ vs $\delta_{min}$, indicating the fatigue limit (continuous line) separating regions in which cyclic crack propagation occurs, or not, for the current parameters ($\bk/\nl=4$, $P=0.5$). 
Within the region of fatigue propagation, the dotted line limits the region in which this propagation can be arrested by a single overload in the system.
Above this line, a region of overload retardation (dashed, not accurately determined) exists.}
\label{phasediagram}
\end{figure}

Whether the combined effect of the larger advance in the overload cycle and reduced advance in ulterior cycles gives an overall retardation or not, depends in a delicate way on the parameters and the values of $\delta$ that have been used. However, one extreme case in which we can be more quantitative is the following. We investigated what the conditions for the largest possible overload to arrest the crack advance are. To answer this question, we first refer to Fig. \ref{phasediagram}. There, the continuous line (picked up from data as that in Fig. \ref{fig:paris_cadenas}) separates the region of cyclic crack propagation (above the curve) from that of no propagation (below). Now, on each situation of a propagating crack, we apply the largest overload the system is able to sustain, namely, one in which $\delta$ is increased to almost $\delta_G^{II}$ 
\footnote{Note that as this overload generates a very large elongation, the result is independent of previous values of $\delta_{min}$ and $\delta_{max}$. This means that a single simmulation of the overload sufficces for all
$\delta_{min}$ and $\delta_{max}$.}. 
This produces a very large elongation during the overload cycle. After that, $\delta$ is reduced to the corresponding $\delta_{min}$, and then increased. The value of $\delta$ at which crack tip starts elongation again (indicated by the dotted line in the figure) is the minimum value of $\delta_{max}$ required for crack elongation after the overload. In other words, in the region between continuous and dotted lines in Fig. \ref{phasediagram}, a crack is arrested by the overload in the system. This quantifies appropriately the conditions for crack arrest in the system, showing clearly that conditions can be found for the process to occur. Typically, above the curve of crack arrest, there is a region of overload retardation (shown shadowed in Fig. \ref{phasediagram}) that  we have not determined in detail for all the parameters. For even larger values of $\delta_{max}$, a net retardation of the crack does not occur, instead the crack gets a net advance due to the overload.

\section{Conclusions}

In this paper we have introduced a simple quasi-one-dimensional model of a crack advancing in between two elasto-plastic chains. The chains 
are joined by breakable springs and attached to two lateral strain gauges by which external loading in the form of prescribed strains are applied to the system. The use of elasto-plastic elements forming the chains gives the possibility of a cyclic fatigue crack advance, in which a finite advance of the crack is obtained at each cycle of the external strain between a maximum and a minimum value. For the case in which we start with a virgin sample (i.e., without plasticity) we have derived analytical expressions to solve the position of the crack tip in terms of the applied strain, and have successfully compared these results with those of numerical simulations. In the case of cyclic advance, we have numerically obtained curves for the amount of elongation in terms of the maximum and minimum strain. We have observed the existence of an endurance limit below which fatigue propagation does not occur, and a maximum strain above which the abrupt rupture of the sample occurs.

The mechanism of fatigue crack advance in the present model is crack tip shielding by a plasticity wake behind the crack tip. It can be described in the following way. During load increase, a stable crack tip advance occurs due to the generation of a plasticity wake behind the crack tip that avoids the immediate unstable propagation of the crack. Upon load reduction, part of the plasticity in the vicinity of the crack tip is reverted. In the second period of load increase, the zone of reverted plasticity allows for a new finite elongation of the crack. As the reverted region is typically smaller than the plastic region induced during crack advance, the elongation 
slightly decreases in consecutive cycles until it reaches a stationary value. The plasticity profile left behind by the advance of the crack 
has typical serrations that reflect in corresponding undulations of the chain profile, reminiscent of the same phenomenon during cyclic fatigue propagation
in real materials.

Crack tip advance per cycle was seen to be a function of both the maximum and minimum value of applied strain $\delta_{max}$ and $\delta_{min}$. This means that it cannot 
be simply written in term of a single $\Delta K$, as in the simplest form of a Paris law. It could  be interpreted that an additional dependence on the stress ratio $R\equiv K_{min}/K_{max}$ exists. 
If a single parameter interpretation of the fatigue advance can be given in our case, this single parameter should be considered to be the amount of plasticity reverted during the unloading part of the cycle. However, this amount depends on 
both $\delta_{max}$ and $\delta_{min}$ giving a two-parameter Paris law.

In addition, we have studied and reproduced the effect of overloading, showing that a cycle with an excess of applied strain can lead to a net retardation in the elongation of the crack, and even to a complete arrest of it. We have numerically determined the conditions in the plane $\delta_{max}$,$\delta_{min}$ for this arrest to occur. The origin of crack retardation or crack arrest in our model is the shielding of the crack tip by the enhanced plastic deformation during the overload cycle.

The minimal model we have presented allows a detailed analysis of the fundamental processes that lead to the possibility of fatigue crack propagation. In a forthcoming publication we plan to discuss the similar properties of a more realistic system consisting of a two-dimensional mesh of elasto-plastic springs and masses loaded in a mode I configuration, where qualitatively the same kind of fatigue propagation is observed.

\section{Acknowledgments}
The authors acknowledge fruitful discussions with Graciela Bertolino and thank Rosario Mar\'{\i}a Schulte for technical assistance . This research was financially supported by Consejo Nacional de Investigaciones Cient\'{\i}ficas y T\'ecnicas (CONICET), Argentina. 

\bibliographystyle{model2-names}

\bibliography{./todaslasreferencias.bib}

\end{document}